\documentclass[aps,prd,reprint,showkeys,nofootinbib,superscriptaddress,notitlepage]{revtex4-1}
\usepackage{amsmath,graphicx,amssymb,multirow,url,xspace}

\usepackage[utf8]{inputenc}
\usepackage[unicode=true,
  bookmarks=false,   backref=false,    colorlinks=true,
  linktocpage=true,  citecolor=black,  linkcolor=black,
  urlcolor=black,    breaklinks=true
]{hyperref}

\newcommand{\GN}{G_{\rm N}}

\newcommand{\hu}{\mathrm{km/s/Mpc}} 

\newcommand{\be}{\begin{equation}}
\newcommand{\ee}{\end{equation}}

\begin{document}

\title{
Analytic Marginalization over Binary Variables in Physics Data
}

\author{Marcus \surname{Högås}}
\email{marcus.hogas@fysik.su.se}
\affiliation{Oskar Klein Centre, Department of Physics, Stockholm University\\Albanova University Center\\ 106 91 Stockholm, Sweden}

\author{Edvard \surname{Mörtsell}}
\email{edvard@fysik.su.se}
\affiliation{Oskar Klein Centre, Department of Physics, Stockholm University\\Albanova University Center\\ 106 91 Stockholm, Sweden}
 
\begin{abstract}
In many data analyses, each measurement may come with a simple yes/no correction—for example, belonging to one of two populations or being contaminated or not.
Ignoring such binary effects may bias the results, while accounting for them explicitly quickly becomes infeasible as each of the $N$ data points introduces an additional parameter, resulting in an exponentially growing number of possible configurations ($2^N$).
We show that, under generic conditions, an exact treatment of these binary corrections leads to a mathematical form identical to the well-known Ising model from statistical physics. This connection opens up a powerful set of tools developed for the Ising model, enabling fast and accurate likelihood calculations. We present efficient approximation schemes with minimal computational cost and demonstrate their effectiveness in applications, including Type Ia supernova calibration, where we show that the uncertainty in host-galaxy mass classification has negligible impact on the inferred value of the Hubble constant.
\end{abstract}

\maketitle

\section{Introduction}
A central task in science is to confront theoretical models with data. 
The likelihood function, which expresses the probability of the observed data given a set of model parameters, is a key ingredient in this process. 
In both Bayesian and frequentist inference, the likelihood underpins parameter estimation, model comparison, forecasting, and simulation. 
Fast and reliable evaluation of the likelihood is hence of fundamental importance.

In many applications, the theoretical prediction for each data point can differ by a discrete offset, for example, when objects belong to one of two populations, when data may or may not be contaminated, or when systematic effects take one of two possible forms. 
Examples include the host-galaxy mass-step correction in Type~Ia supernova calibrations, photometric crowding corrections, or discrete offsets in time-series analyses. 
This introduces $N$ additional binary variables—one for each data point—corresponding to the discrete choice between the two possibilities. 
The resulting growth of the parameter space generically results in methods such as Markov Chain Monte Carlo becoming too costly. 

A natural way forward is to marginalize over the binary choices. 
Here, we show that under broad conditions, this marginalization leads to a correction to the Gaussian likelihood that is formally identical to the partition function of an Ising model, with the binary switches playing the role of spins, the data correlations generating pairwise couplings, the magnitude of the binary offset corresponding to the magnetic moment, and the residuals corresponding to a magnetic field. 
This mapping provides a powerful bridge between statistical data analysis and statistical physics: the Ising model is one of the most thoroughly studied systems in statistical physics. Consequently, an extensive body of exact and approximate techniques have been developed for its computation.

Building on this connection, we present a set of schemes for efficient evaluation of the marginalized likelihood. 
The simplest case—the paramagnetic approximation—adds almost no computational cost beyond that of the baseline Gaussian likelihood, while a more sophisticated approach exploits a mean-field technique. 
After laying out the method in Section~\ref{sec:Method}, we demonstrate its application to concrete physical systems in Section~\ref{sec:Applications}.
Open-source Python implementations of the worked-out examples are available on GitHub.\footnote{\href{https://github.com/marcushogas/Ising-Marginalization}{\url{https://github.com/marcushogas/Ising-Marginalization}}}
\\

\noindent \textbf{Notation.} We will use both index and vector notation. In index form, vectors are written in italics with explicit indices (e.g.\ $y_i$), while in vector form they are written in boldface (e.g.\ $\mathbf{y}$).

\section{Method}
\label{sec:Method}
\subsection{From Binary Shifts to Ising Spins}
Let us assume that we have $N$ observational data points stored in the data vector $y_i^\mathrm{obs}$ and that the model prediction includes a discrete shift like
\begin{equation}
    y_i^{\mathrm{th}}(\theta) \;=\; f_i(\theta) \;+\; s_i\,\Delta_i(\theta),
    \qquad s_i = \pm 1.
\end{equation}
Here, $\theta$ collectively denotes the set of model parameters. The switch $s_i$ toggles an offset $\Delta_i$ up or down. This covers common situations: membership in one of two populations, present/absent contamination, or branch-dependent systematics.

When the binary switches $s_i$ are unknown, we marginalize over all $2^N$ possible assignments. 
The resulting log-likelihood naturally separates into two parts: a ``baseline'' Gaussian term, corresponding to the case with no binary offsets, plus a correction term that accounts for the discrete shifts,
\begin{equation}
\label{eq:LogLikeSplit}
    \ln \mathcal{L} = \ln \overline{\mathcal{L}} + \Delta \ln \mathcal{L}.
\end{equation}
The baseline part is the ordinary Gaussian likelihood describing the fit of $\mathbf{f}(\theta)$ to the data $\mathbf{y}^{\mathrm{obs}}$ and thus reads
\begin{equation}
\label{eq:baseline_log_like}
    \ln \overline{\mathcal{L}} 
    = -\tfrac{1}{2}\!\left(\mathbf{r}^{T} C^{-1}\mathbf{r} + \ln\det C\right),
\end{equation}
where $\mathbf{r}$ is the residual
\begin{equation}
    \mathbf{r}(\theta) \equiv \mathbf{y}^{\mathrm{obs}} - \mathbf{f}(\theta)
\end{equation}
and $C$ is the covariance matrix.
The correction term, $\Delta\ln\mathcal{L}$, contains the statistical weight of the binary switches, and reads,
\begin{equation}
\label{eq:Delta_loglike}
    \Delta \ln \mathcal{L} 
    = \ln \sum_{\mathbf{s}\in\{\pm1\}^{N}}
       \exp\!\left[\tfrac{1}{2}\,\mathbf{s}^{T}J\,\mathbf{s} + \mathbf{s}^{T}\tilde{\mathbf{h}}\right]
       + \tfrac{1}{2}\,\ln\det P
\end{equation}
where we have defined
\begin{subequations}
\label{eq:J_h_defs}
\begin{align}
\label{eq:Jdef}
    J_{ij} &\equiv -\,\Delta_i\,C^{-1}_{ij}\,\Delta_j, \\[3pt]
    \tilde h_i &\equiv h_i + \eta_i, \\[3pt]
    h_i &\equiv \Delta_i \sum_j C^{-1}_{ij}\,r_j, \\[3pt]
    \eta_i &\equiv \tfrac{1}{2}\ln\!\frac{p_i}{1-p_i}, \\[3pt]
    P &\equiv \mathrm{diag}\!\big[p_i(1-p_i)\big].
\end{align}
\end{subequations}
Here, $p_i$ is the prior probability that data point $i$ belongs to the $s_i=+1$ branch.   
The final term in Eq.~\eqref{eq:Delta_loglike}, $\tfrac12\ln\det P$, ensures proper normalization and is only relevant if the probabilities $p_i$ depend on the model parameters themselves.  
A detailed derivation is provided in Appendix~\ref{sec:IsingDerivation}.

Notably, the correction $\Delta \ln \mathcal{L}$ is mathematically identical to the log-partition function of an Ising model with dipoles of spin $s_i=\pm1$ in an effective magnetic field $\tilde{h}_i = h_i + \eta_i$ and pairwise couplings $J_{ij}$, where $\eta_i$ encodes the shift in the magnetic field $h_i$ induced by the prior probability $p_i$ \cite{Ising:1925em,Onsager:1943jn}.
To make the connection complete we may also factorize out the mean variance, $\bar{\sigma}^2$, from the covariance matrix in which case it can be identified as the temperature ($k_B T$). However, for notational simplicity, we refrain from doing so.

\begin{figure}[t]
    \centering
    \includegraphics[width=1\linewidth,trim=0 50 0 0,clip]{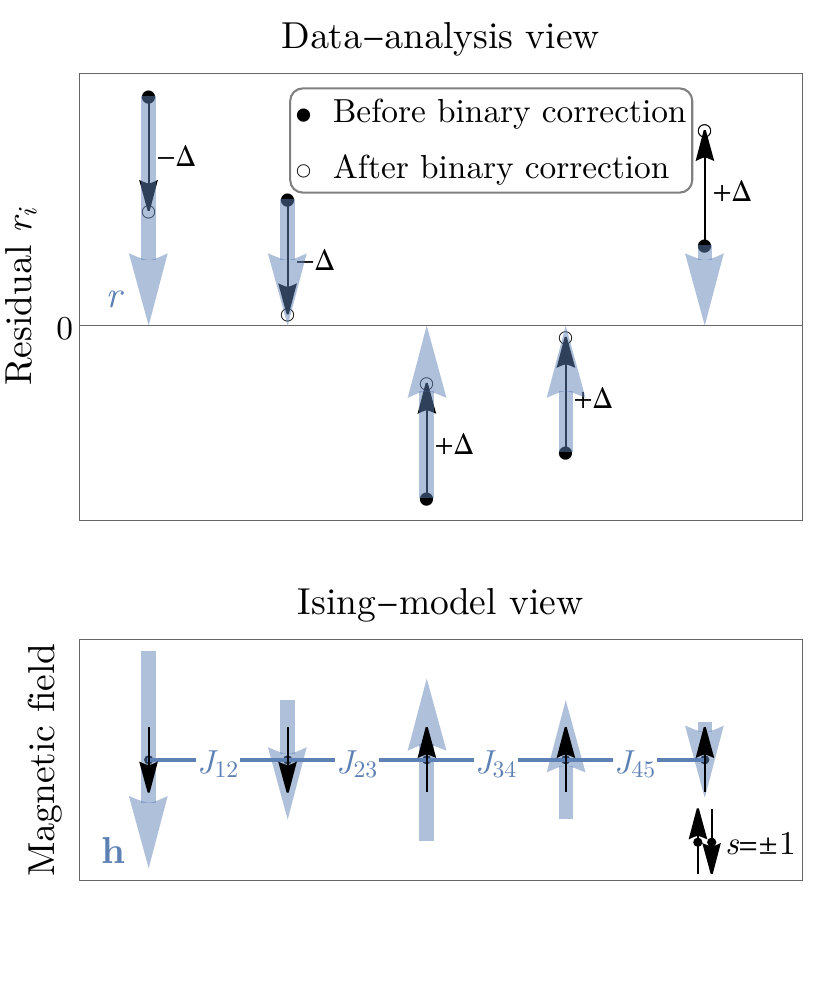}
    \caption{Illustration of the correspondence between the data-analysis and Ising-model views. \emph{Upper panel}: Data-analysis view showing the residuals $r_i$—defined as the difference between data and model before applying the binary corrections $\pm\Delta$—with the residuals indicated by the thick shaded arrows. \emph{Lower panel}: Ising-model view where each data point corresponds to a dipole with spin $s_i = \pm1$ in a magnetic field $\mathbf{h}$. The field strength is proportional to the residual. The strength of the magnetic field shows how decisive the data are—strong fields drive the spins toward one orientation, whereas weak fields correspond to cases where either orientation is nearly equally plausible. Neighbor couplings $J_{ij}$, induced by data correlations, favor aligned ($J_{ij}>0$) or anti-aligned ($J_{ij}<0$) spins. Here, ferromagnetic couplings cause the final dipole to flip against the magnetic field, illustrating how correlations can override an individual data point's preference.}
\label{fig:IsingIllustration}
\end{figure}

A problem that originates in statistical data analysis thus acquires the exact form of a model from statistical physics, bringing with it the full Ising-model toolbox: exact solutions for special graph topologies, mean-field approximations, and more.
It is often helpful to translate between the two perspectives. 
In the data-analysis language, $s_i$ encodes the nature of the binary correction (positive or negative); 
in the statistical-physics language, the same variable is a dipole spin aligned or anti-aligned with the magnetic field $h_i$.
The covariance matrix determines the correlation between the data points. Accordingly, the off-diagonal elements of $C^{-1}_{ij}$ translate directly into the spin--spin couplings $J_{ij}$, cf. Eq.~\eqref{eq:Jdef}. 
Positive off-diagonals $J_{ij}>0$ ($C^{-1}_{ij}<0$) correspond to ferromagnetic interactions, while negative off-diagonals $J_{ij}<0$ ($C^{-1}_{ij}>0$) act as antiferromagnetic ones. 
In the simplified case where correlations are ignored (diagonal $C$), the spins decouple and we have the analogue of a paramagnet.
In this case, the analogue magnetic field $h_i$ reduces to the precision-weighted product of the offset and the residual,
\begin{equation}
    h_i = \frac{\Delta_i}{\sigma_i} \frac{r_i}{\sigma_i}.
\end{equation} 
A magnetic field arises whenever either the shift $\Delta_i$ or the residual $r_i$ is large compared to the corresponding uncertainty, producing a tendency for the dipole to align in one direction.
In data-analysis terms, this corresponds to a strong statistical preference for one offset over the other: when the signal clearly dominates the noise, the selected branch has a decisive impact on the fit quality—manifesting as a large effective magnetic field in the Ising analogy.
Intuitively, the field $h_i$ measures how strongly data point $i$ ``prefers'' one offset over the other.
The correspondence between quantities in the data-analysis problem and those in the Ising model is illustrated in Fig.~\ref{fig:IsingIllustration} and summarized in Tab.~\ref{tab:Dictionary}.  

\begin{table}[t]
  \centering
  \begin{tabular}{l@{\hspace{1.7em}}l}
    \hline \hline
    \textbf{Data analysis} & \textbf{Ising model} \\
    \hline
    Binary switch $s_i = \pm 1$ & Spin up/spin down\\
    Offset $\Delta_i$ & Magnetic moment \\
    Residual $r_i$ & Magnetic field $h_i$ \\
    Prior probability $p_i$ & Shift in the magnetic field $\eta_i$ \\
    Mean variance $\bar{\sigma}^2$ & temperature $k_B T$ \\
    Inverse covariance $C^{-1}$ & Spin--spin interaction $J_{ij}$ \\
    Independent data & Paramagnet \\
    Correlated data & Mean-field approximation$^*$ \\
    \hline \hline
  \end{tabular}
  \caption{Dictionary: Data analysis--Ising model correspondence. 
  When data points have no correlation (diagonal covariance matrix), the likelihood resembles that of a paramagnet (no spin--spin coupling).
  $^*$When data are correlated (non-diagonal covariance matrix), we can use a mean-field approximation to approximate the full interacting Ising model.}
  \label{tab:Dictionary}
\end{table}

With the log-likelihood at hand, the Fisher matrix can be computed to forecast how the binary parameters affect the uncertainties and correlations among the remaining model parameters \cite{Fisher:1935}. This provides a fast way to assess the expected information loss due to the binary uncertainty.
The covariance matrix of the posterior parameter estimates, $\Sigma_{ij}$, is given by the curvature of the log-likelihood as
\begin{equation}
\label{eq:Sigma}
    \Sigma^{-1}_{ij} = - \left. \frac{\partial^2}{\partial \theta_i \partial \theta_j} \left( \ln \overline{\mathcal{L}} + \Delta \ln \mathcal{L} \right) \right|_{\theta = \hat{\theta}},
\end{equation}
where $\hat{\theta}$ is the maximum likelihood estimate (the best fit), see for example Ref.~\cite{Tegmark:1996bz}.
An example calculation is shown in Section~\ref{sec:ToyExample} below.

\subsection{Approximating the Likelihood Correction}
\noindent \textbf{Paramagnetic Approximation.} In practice, we need an efficient way of evaluating $\Delta \ln \mathcal{L}$. The exact form, Eq.~\eqref{eq:Delta_loglike}, is often untractable due to the summation over all spin configurations introducing an $\mathcal{O}(2^N)$ evaluation cost.

A simple approximation can be achieved when ignoring the data point correlations in the correction term $\Delta \ln \mathcal{L}$, that is, ignoring the off-diagonal elements of the covariance matrix in $\Delta \ln \mathcal{L}$.\footnote{To be clear, covariances, if present, are always kept in the baseline likelihood term $\ln \overline{\mathcal{L}}$.}
Unless the data points are strongly correlated, this is a reasonable approximation. In the Ising model, this corresponds to the dipoles being in an external magnetic field, without any pair-wise interactions---a paramagnet. 
In this case, the sum in Eq.~\eqref{eq:Delta_loglike} factorizes and can be expressed analytically as
\begin{equation}
\label{eq:Delta_log_like_Paramg}
    \Delta \ln \mathcal{L} \simeq \frac{1}{2} \mathrm{Tr} \, J + \sum_i \ln \left[ 2 \cosh h_i \right] + \frac{1}{2} \ln \det P.
\end{equation}
An equivalent expression, which is typically more numerically stable in limiting cases such as $p_i \to 0$ or $p_i \to 1$, is
\begin{equation}
\label{eq:Delta_log_like_Paramg2}
    \Delta \ln \mathcal{L} \simeq \frac{1}{2} \mathrm{Tr} \, J + \sum_i \ln \left[ p_i e^{h_i} + (1-p_i) e^{-h_i} \right].
\end{equation}
Since the main computational cost of the baseline Gaussian likelihood lies in inverting the covariance matrix and computing the residual, evaluating the correction term Eq.~\eqref{eq:Delta_log_like_Paramg}/Eq.~\eqref{eq:Delta_log_like_Paramg2} adds virtually no extra cost.\\

\noindent \textbf{Mean-Field Approximation.} To include the effect of correlations, we approximate the likelihood correction $\Delta \ln \mathcal{L}$ using a Hubbard–Stratonovich transformation combined with Laplace’s method. A detailed derivation is presented in Appendix~\ref{sec:ApproxDerivation}. The result is
\begin{equation}
\label{eq:Delta_log_like_nondiag}
    \Delta \ln \mathcal{L} \simeq - \frac{1}{2} \Big( \mathbf{m}^T J \mathbf{m} + \ln \det \left[ D - D^T J D \right] - \ln \det P \Big)
\end{equation}
with $D \equiv \mathrm{diag}(1-m_i^2)$ and $m_i$ is the spin mean-field approximation, subject to the equation
\begin{equation}
\label{eq:MFeqn}
    m_i = \tanh \Big( \tilde{h}_i + \sum_j J_{ij} m_j \Big) .
\end{equation}
This implies that $|m_i| < 1$ and, importantly, there is only one unique solution of Eq.~\eqref{eq:MFeqn}, as shown in Appendix~\ref{sec:ApproxDerivation}.
In the case where $C$, and thus $J$, is diagonal, Eq.~\eqref{eq:Delta_log_like_nondiag} reduces to the paramagnetic approximation, Eq.~\eqref{eq:Delta_log_like_Paramg}, as expected.

Solving the mean-field equation, Eq.~\eqref{eq:MFeqn}, is a potential computational bottleneck. Numerical stability can also be an issue: when the offset if large compared to the uncertainty, $\Delta/\sigma \gtrsim 1$, the steep slope of the $\tanh$ function slows down or prevents convergence, and for strong fields $\tilde{h}_i$ the solutions $m_i$ may approach $\pm 1$, in which case $\ln \det D$ diverges, leading to numerical runaway in Eq.~\eqref{eq:Delta_log_like_nondiag} if not handled carefully. In Section~\ref{sec:Applications} we illustrate these challenges in a concrete application and present practical strategies for how to address them.

\section{Applications}
\label{sec:Applications}
\subsection{Toy Example}
\label{sec:ToyExample}
\begin{figure}[t]
    \centering
    \includegraphics[width=\columnwidth,trim=15 0 40 0,clip]{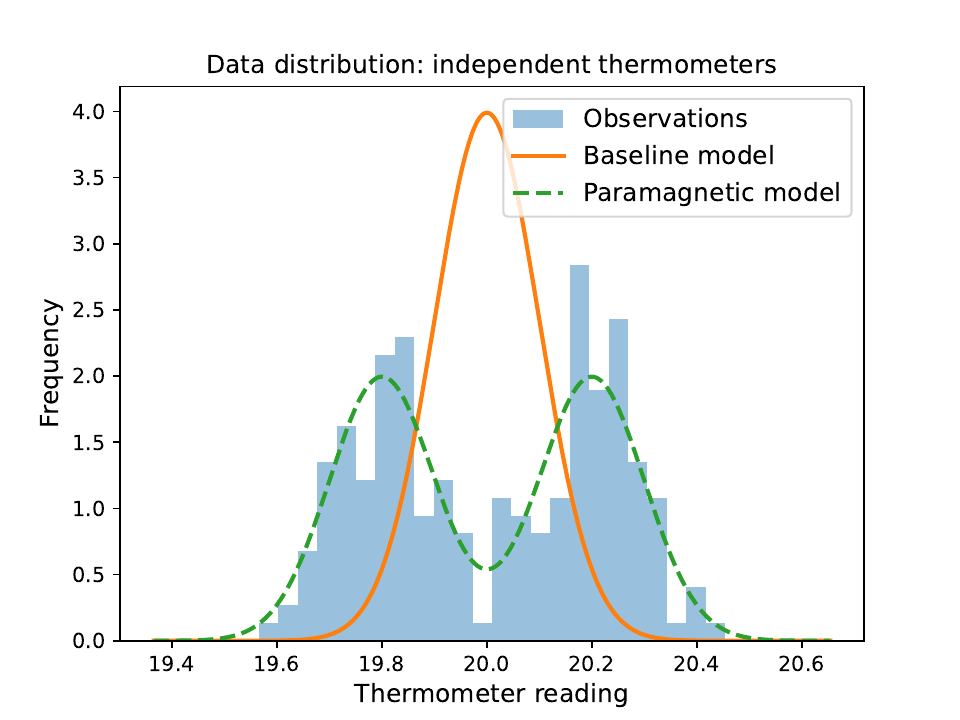}
    \caption{Data distribution for a sample of $N = 200$ thermometers, an offset $\Delta = 0.2$, and measurement uncertainty $\sigma = 0.1$. It is clear that the paramagnetic model including the binary offset, Eq.~\eqref{eq:ToyModel}, is superior compared with the baseline model.}
    \label{fig:data_distr}
\end{figure}

Consider $N$ thermometers measuring a common ambient temperature, $\theta$. Each thermometer may carry a small binary calibration step of known magnitude $\Delta$ (e.g. a factory zero-point shift). The model is
\begin{equation}
\label{eq:ToyModel}
    y_i = \theta + s_i\,\Delta + \varepsilon_i, 
    \quad s_i = \pm 1,
\end{equation}
with a probability $p$ that a given thermometer is positively biased ($s_i = +1$), and, accordingly, a $1-p$ probability for being negatively biased. The measurement noise $\varepsilon_i$ is assumed to be Gaussian with covariance $C$; we study (i) independent thermometers, $C = \mathrm{diag}(\sigma^2)$, and (ii) correlated thermometers, that is, with off-diagonal elements in the covariance matrix. This could for example result from a common environmental effect.
In Fig.~\ref{fig:data_distr} we plot the expected data distribution, comparing the full model of Eq.~\eqref{eq:ToyModel} with the baseline linear model where the binary offsets are neglected.
An open-source Python implementation is available on GitHub.\footnote{ 
\href{https://github.com/marcushogas/Ising-Marginalization}{\url{https://github.com/marcushogas/Ising-Marginalization}}}
\\

\begin{figure}[t]
    \centering
    \includegraphics[width=\columnwidth,trim=0 0 35 0,clip]{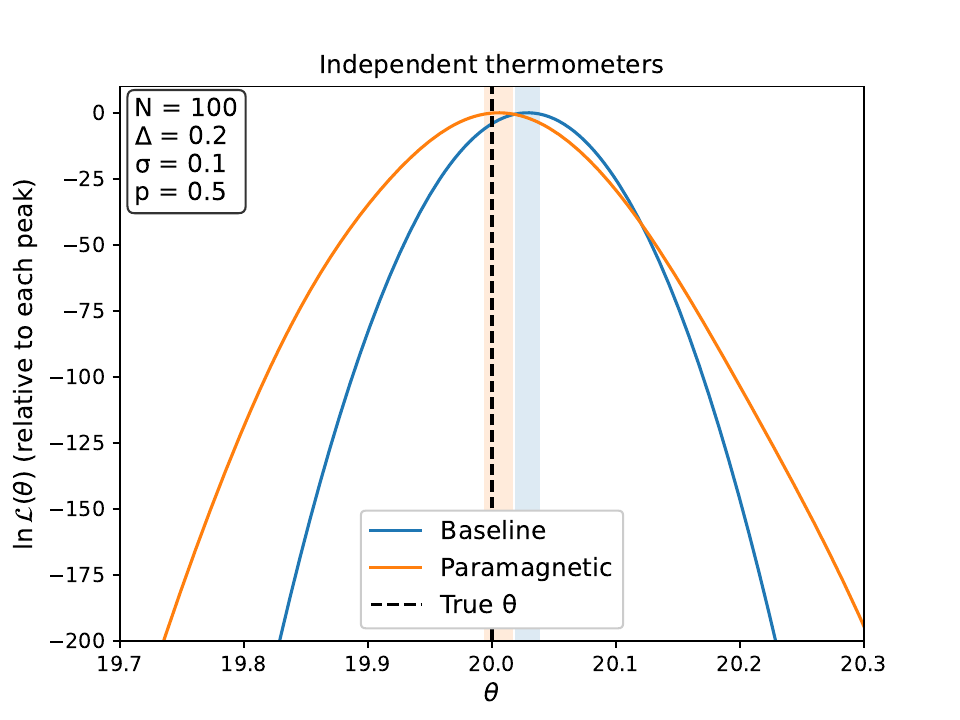}
    \caption{Log-likelihood for the baseline and paramagnetic models. Shaded regions denote the $68 \, \%$ confidence intervals around the best-fit value. Here $\Delta / \sigma = 2$ and the baseline model is biased, here high, with respect to the true value. We also find that the baseline model underestimates the true uncertainty, as it ignores the additional variability introduced by the scatter between positive and negative offsets.}
    \label{fig:indep_thermometers}
\end{figure}

\noindent \textbf{Independent thermometers (paramagnet).} When the covariance matrix is diagonal, there is no correlation between the thermometers and the spins decouple, hence the interaction matrix $J$ is diagonal. In this case $\Delta \ln \mathcal{L}$ assumes the paramagnetic form of Eq.~\eqref{eq:Delta_log_like_Paramg}, and the Fisher matrix simplifies to a single scalar,
\begin{equation}
\label{eq:SigmaToy}
    \Sigma^{-1} = \left. \sum_i \frac{r'_i{}^2}{\sigma^2} \left[ 1 - \left( \frac{\Delta}{\sigma} \mathrm{sech} \, h_i \right)^2  \right] \right|_{\theta = \hat{\theta}}
\end{equation}
where $r'_i = \partial r_i / \partial \theta$. The corresponding uncertainty of the estimated temperature is $\sigma_\theta = \sqrt{\Sigma}$.
The second term inside the bracket quantifies the information loss introduced by the binary parameters, that is, the increase in uncertainty caused by the additional scatter it induces. 
In the example shown in Fig.~\ref{fig:indep_thermometers} the mean value of this correction term is $0.24$, corresponding to a $24 \, \%$ increase in the variance or a $12 \, \%$ increase in the uncertainty $\sigma_\theta$ relative to the baseline case (from $0.010$ to $0.011$). This analytic prediction agrees to excellent precision with the posterior distribution.

\begin{figure}
    \centering
    \includegraphics[width=\columnwidth]{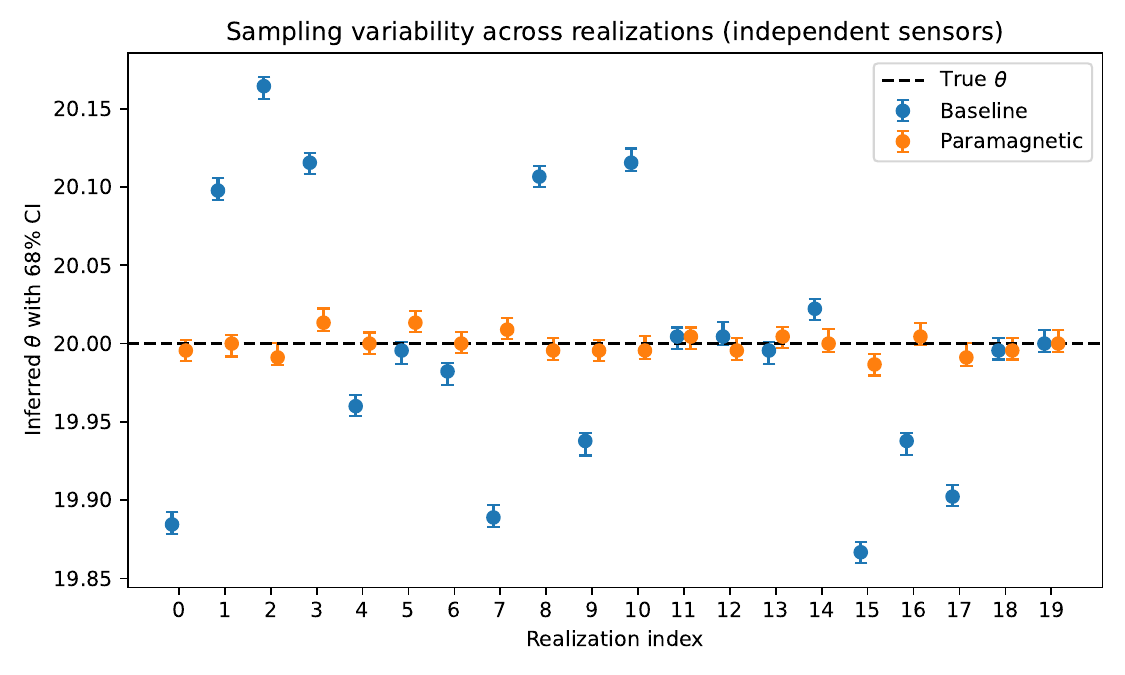}
    \caption{Sample variability of the inferred temperature from the baseline and paramagnetic models. Here, $N = 200$, $\Delta = 1$, $\sigma = 0.1$, and $p = 0.5$. The baseline estimate is biased in the majority of realizations whereas the paramagnetic model consistently reproduces accurate estimates.}
    \label{fig:sampl_variability}
\end{figure}

Even in the simplest case of independent thermometers, the baseline model is biased. The magnitude of this bias increases with the ratio $\Delta/\sigma$ and with any asymmetry in the offset probability ($p \neq 0.5$). Both the estimated mean temperature and its uncertainty are affected, and an accurate inference therefore requires inclusion of the correction term $\Delta \ln \mathcal{L}$.
The sample variability is shown in Fig.~\ref{fig:sampl_variability}.\\

\noindent \textbf{Correlated thermometers (mean-field).} If the thermometers are correlated, the off-diagonal structure of $C^{-1}$ induces spin--spin couplings and the paramagnetic expression is no longer exact. 
Here, we use the mean-field approximation, estimating the likelihood correction using Eqs.~\eqref{eq:Delta_log_like_nondiag}--\eqref{eq:MFeqn}. This introduces some numerical challenges: (i) the evaluation of the log-determinant in Eq.~\eqref{eq:Delta_log_like_nondiag} can become numerically unstable close to $m_i = \pm 1$, leading to runaway behavior, (ii) when the transition in the tanh function is steep (i.e. when $\Delta / \sigma \gtrsim 1$) the mean-field equation Eq.~\eqref{eq:MFeqn} becomes ill-conditioned, slowing down or preventing convergence. 

To address (i), we impose a small floor on the elements of $D$, which prevents floating-point blow-ups as $m_i \to \pm 1$. For (ii), we reparameterize the mean-field equation in terms of $u$, with $m=\tanh u$ and solve it using Newton–Krylov iteration with warm starts, optionally generated through homotopy. These features are implemented in the accompanying Jupyter notebook. While some features therein may be more elaborate than required for this toy example, they are designed to allow seamless extension to more advanced applications.

\begin{figure}
    \centering
    \includegraphics[width=1\columnwidth]{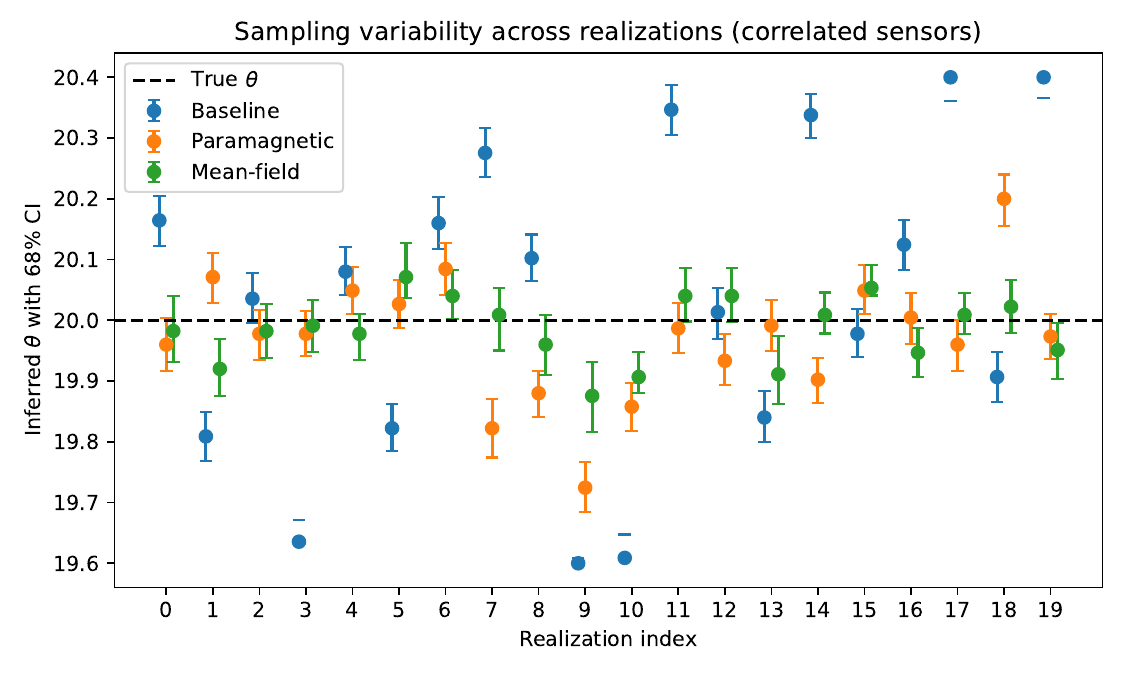}
    \caption{Sample variability of the inferred temperature across 20 realizations with strongly correlated sensors (mean correlation coefficient $\rho \simeq 0.3$). The paramagnetic approximation corrects much of the bias present in the baseline, whereas the mean-field approximation provides a further improvement and is consistent with the true value.}
    \label{fig:samp_var_corr}
\end{figure}

In Fig.~\ref{fig:samp_var_corr} we show the scatter of the inferred temperature across 20 independent data realizations with strong correlations (mean correlation coefficient $\rho \simeq 0.3$ between the data points).

In this toy example we assumed $\Delta$ and $p$ to be known, which is not always the case in realistic settings. However, it is possible to infer both $\Delta$ and $p$ from the data, as demonstrated in the SNe~Ia calibration example below.

\subsection{Type Ia Supernovae Calibration}
Type~Ia supernovae (SNe~Ia) are stellar explosions that serve as standardizable candles that can be used for measuring cosmic distances. Their peak luminosities can be empirically calibrated from the shape (stretch) and color of their light curves, enabling precise distance estimates and, ultimately, measurements of the Hubble constant.
However, several analyses indicate a residual dependence of the standardized brightness on the stellar mass of the host galaxy---first noted in Ref.~\cite{SNLS:2010kps}---where SNe~Ia in high-mass hosts appear slightly brighter, after standardization, than those in low-mass hosts..  
This discontinuity, known as the mass step, is typically modeled as a binary offset in absolute magnitude separated by a threshold around $\log_{10}(M_\star/M_\odot) \simeq 10$.
In this work, we model the mass step as an independent parameter—distinct from color and stretch corrections—in line with Ref.~\cite{Ginolin:2024lfy}. An open-source Python implementation is available on GitHub.\footnote{ 
\href{https://github.com/marcushogas/Ising-Marginalization}{\url{https://github.com/marcushogas/Ising-Marginalization}}}

For supernova~$i$, the standardized apparent magnitude can be written as
\begin{equation}
    m_{B,i} = \mu_i + M_B - \frac{\gamma}{2}\,(1 + s_i),
\end{equation}
with
\begin{equation}
    s_i = 
    \begin{cases}
      +1, & \log_{10} M_i > \log_{10} M_\star,\\[4pt]
      -1, & \log_{10} M_i < \log_{10} M_\star,
    \end{cases}
\end{equation}
and where $\mu_i$ is the distance modulus, $M_B$ the fiducial absolute magnitude, and $\gamma$ the amplitude of the mass step.
In what follows, we fit the observed $B$-band magnitudes, that is, $y_i^\mathrm{obs} = m_{B,i}$.

Equivalently, the magnitude can be expressed in the same form as Eq.~(1), $y_i^{\mathrm{th}}(\theta) = f_i(\theta) + s_i\,\Delta_i$, with
\begin{equation}
    f_i(\theta) = \mu_i + M_B - \frac{\gamma}{2},
    \qquad
    \Delta_i = - \frac{\gamma}{2}.
\end{equation}
In this representation, the mean model prediction $f_i(\theta)$ corresponds to the midpoint between the two populations, while the binary variable $s_i=\pm1$ selects the offset above or below the step.
For simplicity, in the current analysis we treat the supernovae as statistically independent, corresponding to a diagonal covariance matrix $C = \mathrm{diag}( \sigma_{\mathrm{tot},i}^2 )$.

The full supernova sample is divided into two classes: For the calibrators, the distance moduli $\mu_i$ are anchored to independent Cepheid-based distances, and for the Hubble-flow supernovae, the distance moduli follow from the expression
\begin{equation}
\label{eq:muCosmo}
    \mu_i = 5 \log_{10} D_L(z_i) + 25,
\end{equation}
where $D_L(z_i)$ is the luminosity distance in megaparsecs, at redshift $z_i$.
Here, we assume a flat $\Lambda$CDM cosmology, so that $D_L$ depends on the Hubble constant $H_0$ and the matter density parameter $\Omega_m$.

The prior probabilities $p_i$ for supernova $i$ to lie above the host-mass threshold is obtained from the measured host-galaxy masses and their uncertainties as
\begin{equation}
\label{eq:pSN}
    p_i = \int_{\log_{10} M_\star}^{\infty}
    \frac{\exp\!\left[-\tfrac{1}{2}
    \left(\frac{\log_{10} M - \log_{10} M_i}
    {\sigma_{\log_{10} M_i}}\right)^{\!2}\right]}
    {\sqrt{2\pi}\,\sigma_{\log_{10} M_i}}
    \,d\log_{10} M.
\end{equation}
This treatment incorporates the measurement uncertainty of the host mass directly into the Ising framework, in contrast to a sharp deterministic split that assumes each supernova lies definitively on one side of the threshold.

The supernova mass step thus maps directly onto the Ising formulation introduced above.  
Each supernova corresponds to a data point~$i$ with spin~$s_i=\pm1$, the offset $\Delta_i=-\gamma/2$ plays the role of the magnetic moment, and the residuals $r_i$ define an effective magnetic field
\begin{equation}
    h_i = - \frac{\gamma}{2} \frac{r_i}{\sigma_{\mathrm{tot},i}^2}
\end{equation}
which tends to align each spin toward one branch or the other.  
Here, we ignore the correlation between the SNe~Ia and the spin–spin coupling matrix $J_{ij}$ reduces to
\begin{equation}
    J_{ij} = - \left( \frac{\gamma}{2} \right)^2 \frac{\delta_{ij}}{\sigma_{\mathrm{tot},i}^2} .
\end{equation}

\subsubsection*{Generating mock data sets}
To test the method, we generate mock data sets that broadly mimic the statistical properties of the Pantheon+ compilation \cite{Scolnic:2021amr,Brout:2022vxf}, while remaining analytically simple.  
Each mock supernova is assigned the same redshift as in Pantheon+ (after removing duplicate entries).  

To construct the mock dataset, we generate observed distance moduli and standardized magnitudes based on fiducial cosmological and calibration parameters.
For the 42 calibrator supernovae, the observed distance moduli $\mu_i$ are drawn from the Cepheid-based SH0ES values~\cite{Riess:2021jrx}, with Gaussian noise added according to the reported uncertainties.
For the Hubble-flow supernovae, the “observed” moduli are computed from Eq.~\eqref{eq:muCosmo} assuming a flat $\Lambda$CDM cosmology with fiducial parameters $H_0^\mathrm{fid} = 73 \, \hu$ and $\Omega_m^\mathrm{fid} = 0.3$.
The corresponding standardized apparent magnitudes are then simulated as
\begin{equation}
m_{B,i} = \mu_i + M_B^\mathrm{fid} - \frac{\gamma^\mathrm{fid}}{2}(1 + s_i) + \varepsilon_i,
\end{equation}
where $M_B^\mathrm{fid} = -19.253$ and $\gamma^\mathrm{fid} = 0.05 \, \mathrm{mag}$ denote the fiducial absolute magnitude and mass-step amplitude, respectively.
The random noise term~$\varepsilon_i$ is drawn from a Gaussian distribution with $\sigma_B = 0.15 , \mathrm{mag}$, representing the combined photometric and intrinsic scatter.

\begin{figure}[t]
    \centering
    \includegraphics[width=\columnwidth]{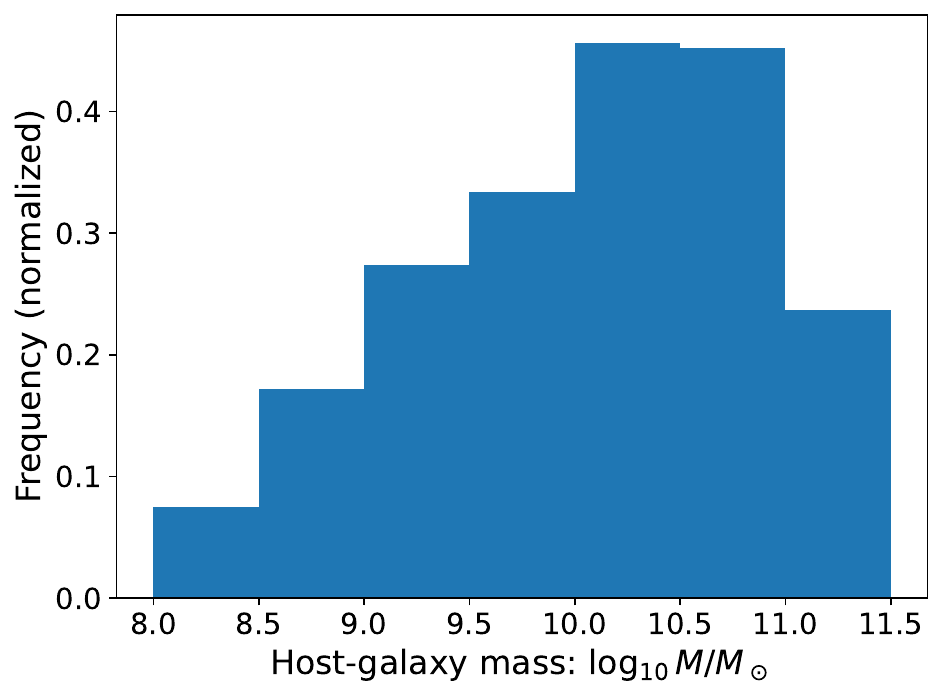}
    \caption{Host-galaxy mass distribution from which the mock dataset masses are sampled.}
    \label{fig:MassDistr}
\end{figure}

\begin{figure*}[t]
    \centering
    \includegraphics[width=1\linewidth]{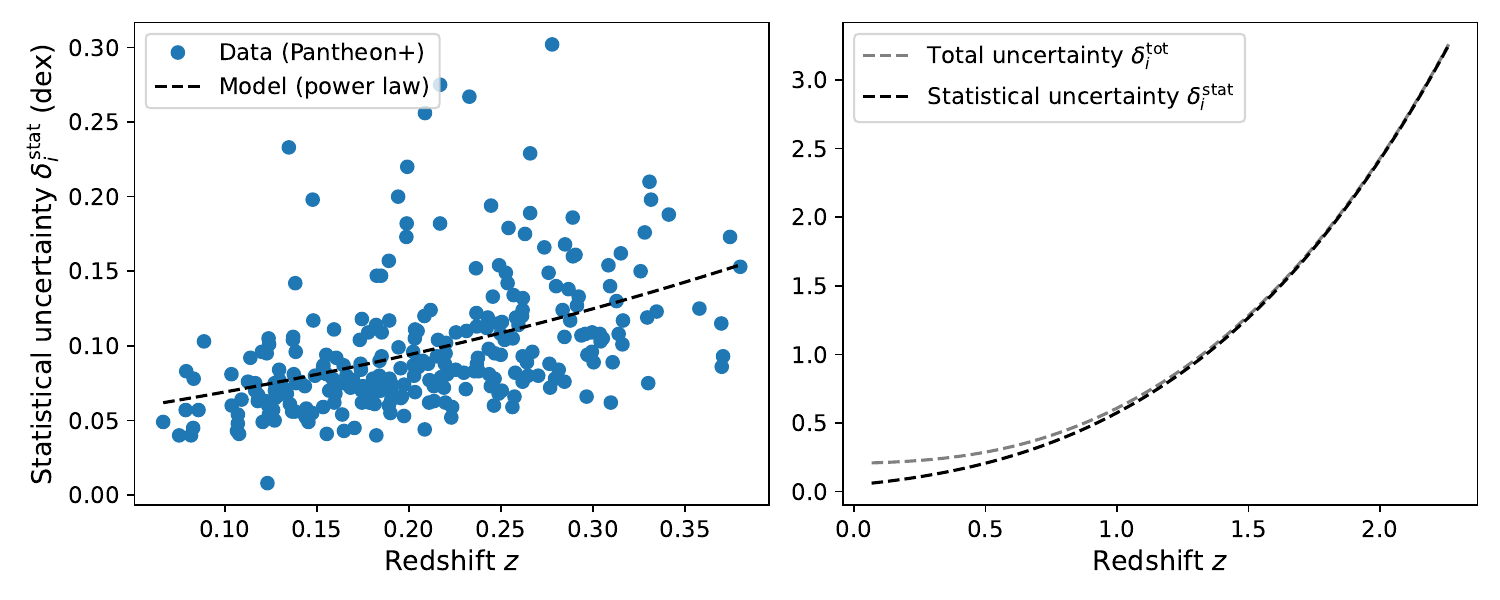}
    \caption{Modeled uncertainty in host-galaxy log-masses. The statistical component is obtained by fitting a power-law model to the host-mass uncertainties tabulated in Pantheon+. A constant method uncertainty of $\delta_\mathrm{meth} = 0.2 \, \mathrm{dex}$, consistent with Pantheon+ \cite{Scolnic:2021amr}, is added in quadrature.}
    \label{fig:logmass_unc}
\end{figure*}

The binary variables $s_i=\pm1$ indicate whether each mock supernova lies above or below the mass threshold $\log_{10} M_\star^\mathrm{fid}/M_\odot=10$ and is drawn with probability $p_i$ according to Eq.~\eqref{eq:pSN}.
For each SN we first draw a fiducial host mass $\log_{10}M_i^{\rm fid}$ from the empirical mass distribution tabulated by Pantheon+ and shown in Fig.~\ref{fig:MassDistr}, preserving the observed fraction of high- and low-mass hosts.
We then draw the (observed) host log-mass by adding a noise term representing the uncertainty in the log-mass estimate,
\begin{equation}
    \log_{10}M_i \sim \mathcal{N}\!\left(\log_{10}M_i^{\rm fid},\,\big[\delta_i^{\rm tot}(z_i)\big]^2\right).
\end{equation}
The total uncertainty $\delta_i^{\rm tot}(z_i)$ is modeled as the quadrature sum of two components:
a redshift-dependent statistical term, $\delta_i^{\rm stat}(z_i)$, which represents the formal uncertainty reported by the host-mass estimation method for each object, and a method uncertainty, $\delta_{\rm meth}$, capturing the systematic inter-method variation in host-mass estimates—i.e., the intrinsic precision limit of the technique,
\begin{equation}
    \delta_i^{\rm tot}(z_i)=\sqrt{\big[\delta_i^{\rm stat}(z_i)\big]^2+\delta_{\rm meth}^2}.
\end{equation}
The redshift dependence in the statistical uncertainty is modeled as a power law,
\begin{equation}
    \delta_i^{\rm stat}(z_i)=c\,(1+z_i)^n,
\end{equation}
where the parameters $c$ and $n$ are determined from a best fit to the tabulated host-mass uncertainties in the Pantheon+ sample, see Fig.~\ref{fig:logmass_unc}.

We generate two mock data sets that differ only in their redshift selection and refer to these as the ``SH0ES selection'' and the ``high-$z$ selection.''
The first follows the SH0ES sample, including the Cepheid calibrator supernovae together with the subset of Hubble-flow events used in the SH0ES $H_0$ calibration.  
The second uses a broader selection, containing all supernovae with $z > 0.023$ in addition to the calibrators, corresponding to the full Pantheon+ redshift range.  
The low-redshift cutoff minimizes the impact of peculiar velocities on the inferred distances.

This mock data generation reproduces the key features of the Pantheon+ data set with realistic measurement noise, and host-mass distribution—while retaining full control over the true underlying parameters.
Unlike the real Pantheon+ sample, however, we neglect correlations between data points and assume a diagonal covariance matrix with a common uncertainty in the $B$-band magnitude of $\sigma_B = 0.15 \, \mathrm{mag}$ for all supernovae.  
For the calibrator subsample, the total uncertainty is augmented by the corresponding uncertainty in the Cepheid-based distance modulus, added in quadrature, $\sigma^2_{\mathrm{tot},i} = \sigma_B^2 + \sigma_{\mu,i}^2$.

This approach provides a clean environment for validating the Ising-based marginalization method.

\subsubsection*{Results}
\begin{figure*}[t]
    \centering
    \includegraphics[width=1\linewidth]{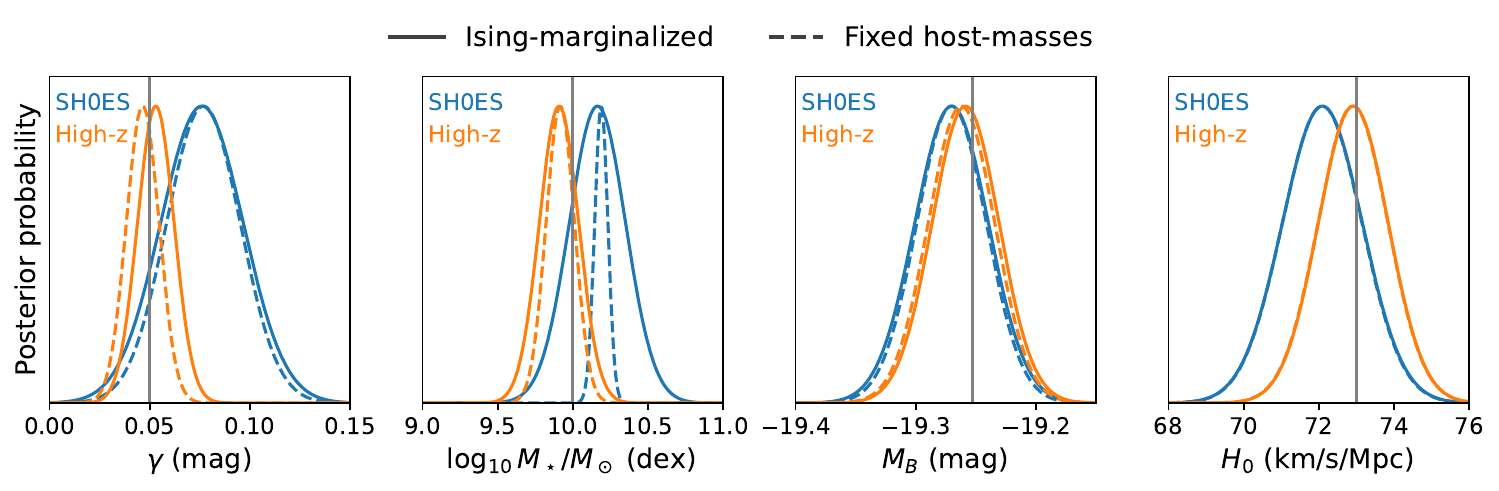}
    \caption{1D marginalized posterior distributions of the mass-step amplitude $\gamma$, the host-mass threshold $\log_{10} M_\star$, the absolute magnitude $M_B$, and the Hubble constant $H_0$. Blue curves correspond to the SH0ES-like sample and orange curves to the extended high-$z$ sample. Solid lines show the Ising-marginalized results, which account for uncertainties in the host-galaxy masses, while dashed lines show the traditional approach, where each supernova is assigned deterministically to one side of the mass step. The vertical gray lines denote the fiducial parameter values. The Ising method accurately recovers all fiducial parameters, whereas the traditional approach yields underestimated uncertainties of the mass step.}
    \label{fig:1Dmarg}
\end{figure*}

Using mock datasets constructed as described above, we fit the standardized $B$-band magnitudes with the Ising-marginalized likelihood to obtain posterior distributions for the model parameters.  
Examples are shown in Fig.~\ref{fig:1Dmarg}.  
As expected, the precision of the inferred parameters improves with sample size, with the broader, high-$z$, dataset yielding narrower posteriors for both the astrophysical and cosmological parameters.

To illustrate the influence of the host-mass uncertainty, as accounted for by the Ising method, we also perform fits in which the uncertainties in $\log_{10} M$ are ignored—that is, each supernova is assigned to one side of the mass threshold based solely on its best-estimate host mass.  
This corresponds to the traditional treatment of the SN~Ia mass step in current analyses.  
As shown in Fig.~\ref{fig:1Dmarg}, the conventional approach reproduces the best-fit values reasonably well but systematically underestimates the uncertainties in the inferred mass step. A clear example is the inferred mass-step location for the SH0ES-like dataset: the Ising method yields $\log_{10}(M_\star / M_\odot) = 10.16 \pm 0.19 \, \mathrm{dex}$, consistent with the fiducial value within one standard deviation, whereas the fixed-mass approach gives $\log_{10}(M_\star / M_\odot) = 10.19 \pm 0.04 \, \mathrm{dex}$, five standard deviations away from the fiducial value.
This demonstrates that the Ising-based marginalization correctly propagates the additional uncertainty from the binary population assignments into the posterior distribution which is not the case in the traditional approach.

While the uncertainty in the inferred mass step is strongly affected by the host-galaxy mass uncertainty, the calibration of the absolute magnitude $M_B$ and the cosmological parameters ($H_0$ and $\Omega_m$) remain largely unchanged, as shown in Fig.~\ref{fig:1Dmarg}.
A Fisher-information analysis, Eq.~\eqref{eq:Sigma}, confirms these trends quantitatively with
\begin{equation}
\label{eq:SigmaH0H0}
    \Sigma_{H_0 H_0}^{-1} = \sum_i \frac{r'_i{}^2}{\sigma_{\mathrm{tot},i}^2} \left[ 1 - \left( \frac{\gamma}{\sigma_{\mathrm{tot},i}} \right)^2 \frac{p_i (1 - p_i)}{A_i^2} \right]
\end{equation}
where $r'_i = \partial r_i / \partial H_0$ and 
\begin{equation}
    A_i \equiv p_i e^{h_i} + (1-p_i) e^{-h_i}.
\end{equation}
The curvature term in the residual, $\partial^2 r_i / \partial H_0^2$, has been neglected.
This omission is standard in Fisher analyses, as the second-derivative term typically contributes only weakly near the likelihood maximum where the residuals are small and the log-likelihood is well approximated by a local quadratic expansion.

Equation~\eqref{eq:SigmaH0H0} makes explicit how the mass step affects the Fisher information for $H_0$: it enters through the second term inside the square brackets. The factor $p_i(1-p_i)/A_i^2$ quantifies the information loss associated with the uncertainty in the mass step.
For a specific supernova, it reaches its maximum when its mass-step assignment is maximally ambiguous—namely when the prior probability is completely uncertain ($p_i = 0.5$) and the data provide no preference ($h_i = 0$). In this case, the factor evaluates to $\max p_i(1-p_i)/A_i^2 = 1/4$.
With $\gamma = 0.05 \, \mathrm{mag}$ and $\sigma_{\mathrm{tot},i} \geq 0.15 \, \mathrm{mag}$, the mass step can at most reduce the Fisher information for $H_0$ by $2.8 \, \%$.
In practice, the effect is much smaller, since the majority of SNe~Ia are confidently classified on one side of the mass threshold or the other (over $90 \, \%$ in our mock samples).
Therefore, we expect a negligible contribution to the uncertainty in $H_0$ from the mass step, which is confirmed by the results presented in Fig.~\ref{fig:1Dmarg}.

\section{Discussion}
While our analysis of the supernova mass step shows that, under ideal conditions, the uncertainty in the host-galaxy mass estimates has negligible impact on the inferred value of $H_0$, this conclusion relies on the assumption that the underlying mass distributions of the calibrator and Hubble-flow samples are free from systematic offsets.
For instance, if the stellar masses of the calibrator hosts are systematically overestimated relative to those of the Hubble-flow galaxies, some calibrators will be incorrectly classified as high-mass hosts.
In this case, the mass-step correction applied during standardization makes the calibrator supernovae appear intrinsically fainter (larger $M_B$) than they truly are.
When this fainter zero point is applied to the Hubble-flow supernovae, their inferred distances become systematically smaller than their true values, leading to an overestimation of the Hubble constant.
The Ising-marginalization method addresses random uncertainty in $\log_{10} M$ but cannot by itself correct for coherent systematic shifts between samples. Therefore, while we have shown that the stochastic uncertainty in host mass has little effect on cosmological inference, systematic differences in mass estimation between calibration rungs remain a potentially important source of bias and must be carefully monitored.

Another application arises in the calibration of the first rung of the distance ladder, where classical Cepheids serve as standard candles.
Cepheids occur in distinct pulsation modes—primarily the fundamental (F) and first overtone (1O)—which follow offset period–luminosity (PL) relations and, in some bands, exhibit slightly different slopes.
For instance, measurements in the Large Magellanic Cloud (LMC) show that the $W_I$-band PL relations of fundamental and first-overtone Cepheids differ both in zero point and slope, with the 1O sequence offset toward fainter magnitudes at fixed period \cite{Soszynski:2008kd}.
In current distance-ladder calibrations, including that of the SH0ES team, overtone contamination is minimized by applying period cuts (typically $P > 5~\mathrm{d}$), effectively removing the majority of 1O Cepheids \cite{Riess:2021jrx}.

The Ising-marginalization framework offers an alternative: to include both modes simultaneously while marginalizing over their uncertain classification.
As a simple example, for Cepheid $i$ with period $P_i$, one may write the standardized (e.g., Wesenheit) magnitude model as
\begin{equation}
m_i^{\mathrm{th}} = \alpha \log_{10} P_i + \beta + s_i \, \Delta_i,
\qquad s_i = \pm 1,
\end{equation}
where $s_i = +1$ denotes a fundamental-mode Cepheid and $s_i = -1$ a first-overtone Cepheid.
The offset $\Delta_i$ captures the mode-dependent magnitude shift, which may also include a weak period dependence, $\Delta_i = \Delta_0 + \Delta_1 (\log P_i - \log P_0)$, if the two PL relations differ in slope.
This formulation maps directly onto the Ising formalism, with the pulsation mode $s_i$ being analogous to a spin orientation, the corresponding magnitude offset $\Delta_i$ being analogous to a magnetic moment, and correlations between Cepheids translate into spin--spin couplings $J_{ij}$.
The prior probability $p_i$ for each mode can for example be constructed from light-curve–shape diagnostics such as Fourier parameters ($R_{21}$, $\phi_{21}$).

Since the Ising marginalization correctly propagates the uncertainty in the pulsation-mode classification, this approach potentially allows short-period Cepheids to be included in the distance-ladder calibration, improving statistical precision while maintaining an accurate treatment of classification uncertainty.

Another application of the Ising-marginalization framework is on fifth-force effects on Cepheid luminosities.
In screened modified gravity models, the local gravitational strength can vary as $G = \GN + \Delta G$, which alters stellar structure and shifts the Cepheid PL relation \cite{Sakstein:2019qgn,Desmond_2019,Hogas:2023pjz}.
The magnitude of this shift depends, among other things, on whether a Cepheid lies on the second or third crossing of the instability strip—two distinct evolutionary phases.
In principle, a Cepheid’s instability-strip crossing number can be inferred from secular period changes, that is, the sign and magnitude of $\dot{P}$. In practice however, this relies on extensive, long-baseline, and well-sampled datasets.
In Ref.~\cite{Turner:2006pi}, evolutionary trends in the Cepheid periods became detectable only after inclusion of archival photographic plate data, extending the observational baseline to several decades.
Such decades-long monitoring is not available for the Hubble Space Telescope Cepheids employed in the calibration of the cosmic distance ladder.
Using the Ising method however allows for the inclusion of this uncertainty in a statistically consistent way, automatically propagating it into the calibration of the distance ladder and the inferred value of $H_0$ \cite{Hogas:2025}.\footnote{In the standard calibration (no fifth force) the instability-strip crossing number plays only a minor role, affecting the mean luminosity by at most a few hundredths of a dex \cite{2024A&ARv..32....4B}.}

\section{Conclusions}
We have introduced the Ising-marginalization framework---an analytic method for marginalizing over binary nuisance variables that appear in statistical data analyses.  
By showing that this marginalization leads to a log-likelihood correction formally identical to the partition function of an Ising model, we have established a direct link between statistical data analysis and statistical physics.  
This correspondence provides both conceptual insight and access to a rich set of approximation techniques, including the efficient paramagnetic and mean-field schemes presented here.

Using two examples, we have validated the method and demonstrated its efficiency.  
In a simple toy model, the Ising approach accurately recovers the underlying parameters while properly inflating the inferred uncertainties.  
Applied to Type~Ia supernova calibration, it accurately reproduces the fiducial mass-step parameters and shows that the uncertainty in the host-galaxy mass classification has negligible impact on the inferred value of the Hubble constant.  

We have also outlined potential extensions of the method to other rungs of the distance ladder, including the treatment of Cepheid overtone classification and the crossing-number ambiguity relevant for fifth-force tests in modified gravity models.  
These examples demonstrate the flexibility of the approach: any binary source of uncertainty can be incorporated in this framework without inflating the dimensionality of the parameter space.

Although our focus has been on astrophysical calibration, the method is fully general.  
It can be applied to a broad class of inference problems across physics and beyond, wherever discrete uncertainties or classification ambiguities limit the fidelity of statistical modeling.

\begin{acknowledgements}
We thank D’Arcy Kenworthy for suggesting the application of the Ising-marginalization framework to Cepheid overtone classification.

OpenAI's ChatGPT has been used to draft portions of the manuscript.

MH and EM acknowledges support from the Swedish Research Council under Dnr VR 2024-03927.
\end{acknowledgements}

\section*{Data availability}
The Pantheon+ data is publicly available at \url{https://github.com/PantheonPlusSH0ES/DataRelease/tree/main/Pantheon%2B_Data}.

The Python implementation of the Ising-marginalization method, including the toy example and the Type~Ia supernova mass-step calibration, is publicly available on GitHub at
\href{https://github.com/marcushogas/Ising-Marginalization}{\url{https://github.com/marcushogas/Ising-Marginalization}}.

\appendix
\section{Deriving the Ising model}
\label{sec:IsingDerivation}
Assuming Gaussian uncertainties and a covariance matrix $C$, possibly dependent on the model parameters $\theta$, we can write the likelihood
\begin{equation}
    \mathcal{L} = \frac{ \exp \left[ - \frac{1}{2} (\mathbf{y}^\mathrm{obs} - \mathbf{y}^\mathrm{th})^T \, C^{-1} \, (\mathbf{y}^\mathrm{obs} - \mathbf{y}^\mathrm{th}) \right] }{\sqrt{\det C}} .
\end{equation}
Defining the residual between the observations and the mean model prediction as
\begin{equation}
    r_i(\theta) = y_i^\mathrm{obs} - f_i(\theta),
\end{equation}
the likelihood can be rewritten as
\begin{equation}
    \mathcal{L} \propto \frac{\exp\left[ -\frac{1}{2} \mathbf{r}^T \, C^{-1} \, \mathbf{r} \right] \exp\!\left[
    \tfrac{1}{2} \mathbf{s}^{T} J\, \mathbf{s} + \mathbf{s}^{T}\mathbf{h} \right]}{\sqrt{\det C}} 
\end{equation}
where we have defined
\begin{equation}
    h_i \equiv \Delta_i \sum_j C^{-1}_{ij} r_j, \quad
    J_{ij} \equiv - \Delta_i C^{-1}_{ij} \Delta_j.
\end{equation}
Therefore, the total log-likelihood can be split into a ``baseline'' part plus a correction,
\begin{equation}
    \ln \mathcal{L}(\theta) = \ln \overline{\mathcal{L}}(\theta) + \Delta \ln \mathcal{L}(\theta)
\end{equation}
where
\begin{subequations}
    \begin{align}
        \ln \overline{\mathcal{L}}(\theta) &\equiv - \frac{1}{2} \left( \mathbf{r}^T C^{-1} \mathbf{r} + \ln \det C \right), \\
        \Delta \ln \mathcal{L}(\theta) &\equiv   \frac{1}{2} \mathbf{s}^{T} J\, \mathbf{s} + \mathbf{s}^{T}\mathbf{h} .
    \end{align}
\end{subequations}
The baseline, $ \ln \overline{\mathcal{L}}(\theta)$, is the log-likelihood in absence of any shift ($\Delta_i = 0$).
When the binary switches $s_i$ are unknown, we marginalize by summing the likelihood over all possible assignments of $s_i$. Let $p_i$ denote the a prior probability of $s_i$ being in the state $+1$ (thus $1-p_i$ is the probability of being in state $-1$), then the marginalized likelihood reads
\begin{widetext}
    \begin{equation}
    \Delta \ln \mathcal{L}(\theta) = \ln \sum_{\mathbf{s} \in \{ \pm 1 \}^N } \left[ \left( \prod_i p_i^{ (1+s_i)/2 } (1-p_i)^{ (1-s_i)/2 } \right) \exp \left[ \tfrac{1}{2} \mathbf{s}^{T} J\, \mathbf{s} + \mathbf{s}^{T}\mathbf{h} \right] \right].
\end{equation}
\end{widetext}
The factor mutiplying the exponent represents the \emph{a priori} probability of that particular configuration of offsets, $s_i$. If we put the probabilities in the exponent (using \mbox{$p^n = e^{n \ln p}$}), the product becomes a sum and we get
\begin{widetext}
    \begin{equation}
        \Delta \ln \mathcal{L}(\theta) = \ln \sum_{\mathbf{s} \in \{ \pm 1 \}^N} \exp \left[ \frac{1}{2} \mathbf{s}^{T} J\, \mathbf{s} + \mathbf{s}^{T}\mathbf{h} + \sum_i \frac{s_i}{2} \ln \frac{p_i}{1-p_i} + \sum_i \frac{1}{2} \ln p_i (1-p_i) \right] .
    \end{equation}
\end{widetext}
The last term in the exponent can be factorized out of the sum, leading to the normalization term seen in Eq.~\eqref{eq:Delta_loglike}.
The parts which are linear in $s_i$ in the exponent can then be merged into a single term if we define a shifted magnetic field
\begin{equation}
    \tilde{h}_i = h_i + \eta_i
\end{equation}
with
\begin{equation}
    \eta_i \equiv \frac{1}{2} 
    \ln \!\ \frac{ p_i }{1 - p_i}.
\end{equation}
In other words, if the prior probability of spin up ($s_i = +1$) is greater than $50 \, \%$ ($p_i > 0.5$) the effective magnetic field $\tilde{h}_i$ acquires a positive shift, effectively increasing its strength in the positive (up) direction.
Since the dipole wants to align with the magnetic field, this results in an increased tendency of spin up, as expected.

It follows that
\begin{equation}
    \Delta \ln \mathcal{L} 
    = \ln \sum_{\mathbf{s}\in\{\pm1\}^{N}}
       \exp\!\left[\tfrac{1}{2}\,\mathbf{s}^{T}J\,\mathbf{s} + \mathbf{s}^{T}\tilde{\mathbf{h}}\right]
       + \tfrac{1}{2}\,\ln\det P,
\end{equation}
which completes the derivation of Eq.~\eqref{eq:Delta_loglike}.
The derivation can also be generalized to the case where the prior probability $p_i$ itself follows a prior probability distribution $\pi_i(p_i)$, for example if we have some prior knowledge of the probabilities.  
In this case we can marginalize also over $p_i$ resulting in a set of integrals of the following form
\begin{widetext}
    \begin{equation}
    \int_0^1 dp_i \pi_i(p_i) p_i^{(1+s_i)/2} (1-p_i)^{(1-s_i)/2} = \left\lbrace \begin{array}{ll}
        \left\langle p_i \right\rangle, \quad & \mathrm{if} \; s_i = +1 \\
        1 - \left\langle p_i \right\rangle, \quad & \mathrm{if} \; s_i = -1
    \end{array} \right.
\end{equation}
\end{widetext}
with $\langle p_i \rangle$ being the expectation value of $p_i$,
\begin{equation}
    \langle p_i \rangle \equiv \int_0^1 dp_i \, \pi_i(p_i)\, p_i.
\end{equation}
Hence, the derivation above remains valid even when $p_i$ has its own prior distribution, provided that $p_i$ is replaced by its prior expectation value.

\section{Deriving the Mean-Field Approximation}
\label{sec:ApproxDerivation}
Here, we derive an approximation of the likelihood correction (ignoring the normalization in $p$, which can simply be added at the end),
\begin{equation}
\label{eq:Delta_log_like_nondiag_2}
    \Delta \ln \mathcal{L} = \ln \sum_{ \mathbf{s} \in \{ \pm 1 \}^N } \exp \left[ \tfrac{1}{2} \mathbf{s}^T J \mathbf{s} + \mathbf{s}^T \mathbf{\tilde{h}} \right]
\end{equation}
including the effect of non-diagonal components of $J$ ($C$). 
As a first step, we rewrite Eq.~\eqref{eq:Delta_log_like_nondiag_2} using a Hubbard--Stratonovich transform, converting the discrete particle problem $s_i = \pm 1$ into a field theory $\phi_i \in \mathbb{R}^N$:
\begin{widetext}
    \begin{equation}
        \tfrac{1}{2} \mathbf{s}^T J \mathbf{s} = (2 \pi)^{-N/2} \left( \det (-J) \right)^{-1/2} \int_{ \mathbb{R}^N } d^N\phi \, \exp \left[ \tfrac{1}{2} \sum_{i,j} \phi_i J^{-1}_{ij} \phi_j + \mathrm{i} \, \sum_i \phi_i s_i \right] .
    \end{equation}
\end{widetext}
Using this relation, we obtain
\begin{widetext}
    \begin{equation}
        \Delta \ln \mathcal{L} = \left( \det (-J) \right)^{-1/2} \int_{ \mathbb{R}^N } d^N\phi \, \exp \left[ \tfrac{1}{2} \sum_{i,j} \phi_i J^{-1}_{ij} \phi_j \right] \sum_{ \mathbf{s} \in \{ \pm 1 \}^N } \exp \left[ \sum_i s_i (\tilde{h}_i + \mathrm{i} \, \phi_i) \right]
    \end{equation}
\end{widetext}
which can be rewritten as
\begin{equation}
\label{eq:ExpInt}
    \Delta \ln \mathcal{L} = \left( \det (-J) \right)^{-1/2} \int_{ \mathbb{R}^N } d^N\phi \, e^{S(\phi)}
\end{equation}
with
\begin{equation}
    S(\phi) \equiv \tfrac{1}{2} \sum_{i,j} \phi_i J^{-1}_{ij} \phi_j + \sum_i \ln \left[ 2 \cosh (\tilde{h}_i + \mathrm{i} \, \phi_i) \right].
\end{equation}
The integral \eqref{eq:ExpInt} can be approximated using Laplace's method, expanding the exponent around its global maximum. 
The stationary point (the maximum) is situated at root of the gradient, that is, where $\partial S(\phi) / \partial \phi_i = 0$. Interestingly, this yields the mean-field equation 
\begin{equation}
\label{eq:MFeqn2}
    m_i = \tanh \Big( \tilde{h}_i + \sum_j J_{ij} m_j \Big)
\end{equation}
where we have defined
\begin{equation}
    m_i(\phi) \equiv \tanh (\tilde{h}_i + \mathrm{i} \, \phi_i) .
\end{equation}
A Taylor expansion around the stationary point, $\phi_\star$, yields
\begin{equation}
\label{eq:Taylor_S}
    S(\phi) \simeq S(\phi^\star) + \tfrac{1}{2} \sum_{i,j} (\phi_i - \phi_i^\star) H_{ij} (\phi_j - \phi_j^\star)
\end{equation}
where $H_{ij}$ is the Hessian matrix,
\begin{equation}
    H_{ij} \equiv \left. \frac{\partial^2 S(\phi)}{\partial \phi_i \partial \phi_j} \right|_{\phi^\star} .
\end{equation}
The Hessian matrix can be written
\begin{equation}
    H = J^{-1} - D, \quad D \equiv \mathrm{diag}(1-m_i^2).
\end{equation}
With these results, the integral \eqref{eq:ExpInt} can be approximated using Eq.~\eqref{eq:Taylor_S} with the result
\begin{equation}
    \Delta \ln \mathcal{L} \simeq -\tfrac{1}{2} \ln \det (-J) - \tfrac{1}{2} \ln \det (-H) + S(\phi^\star).
\end{equation}
After some algebraic simplifications we finally get
\begin{equation}
    \Delta \ln \mathcal{L} \simeq - \frac{1}{2} \left( \mathbf{m}^T J \mathbf{m} + \ln \det \left[ D - D^T J D \right] \right).
\end{equation}
To get the final expression for $\Delta \ln \mathcal{L}$ of Eq.~\eqref{eq:Delta_log_like_nondiag}, we add the normalization term in $p_i$.
The only remaining question is that of the uniqueness of the solution to the mean-field equation \eqref{eq:MFeqn2}.
The ordinary Ising model exhibits one or three solutions, depending on the temperature. 
However, here the Hessian $H = J^{-1} - D$ is a symmetric, negative definite matrix for all $|m_i| < 1$ ($J^{-1}$ is symmetric, negative semi-definite). Hence, $S(\phi)$ is strictly concave and the stationary point $\phi^\star$ must therefore be unique, as well as the solution of Eq.~\eqref{eq:MFeqn2}.

\clearpage
\bibliography{bibliography}{}
\bibliographystyle{apsrev4-1}

\end{document}